\documentclass[letterpaper]{jpconf}
\usepackage{graphicx}
\usepackage{epstopdf}

\begin{document}

\title{Forward Physics and BRAHMS results}

\author{Ramiro Debbe\dag\ for the BRAHMS Collaboration  }

\address{\dag\ Brookhaven National Laboratory, Upton NY, 11973 }

\begin{abstract}
We report here the BRAHMS measurements of particle production in d+Au and p+p collisions at RHIC. The results
presented here are compared to previous p+A measurements at lower energies in fixed target mode. Some preliminary
results on abundances of identified particles at high rapidity are also presented.

\end{abstract}




\section{Introduction}
The first collisions of gold ions at nucleon-nucleon center of mass energies 
$\sqrt{s_{NN}}=130 GeV$ at RHIC showed a dramatic drop in the production of
pions at intermediate $p_{T}$ compared to an incoherent sum of pions produced in  p+p collisions 
at the same energy (approximately a factor of 5 for the most central 
collisions) \cite{PHENIX}.
The measured suppression could be the result of energy degradation of jets 
traversing a newly formed dense medium or could also be related to 
modifications to the wave function of the ions brought in by the high energy 
of the collisions together with the high atomic number A of the nuclei \cite{KLM}. Collisions between deuterium 
and gold ions at $\sqrt{s_{NN}}=200 GeV$ 
were produced during the third RHIC run to resolve the apparent conflict 
between the above mentioned explanations of the suppression.
Particle production from d+Au collisions around mid-rapidity do not show the suppression seen in Au+Au 
collisions \cite{RHICdA, BRAHMS-Supp}. What is seen instead is an
enhancement that has been associated with the so called ``Cronin effect'' 
\cite{Antresyan}, where partons undergo multiple incoherent scatterings that
increase their transverse momenta as they traverse through the target.
These results constitute evidence that a new dense and highly opaque medium 
has been formed in Au+Au collisions at RHIC \cite{WhitePapers} and the suppression of intermediate to 
high $p_{T}$ leading particles is directly related to
their interaction with that medium be it collisional or by induced gluon 
radiation.

But the unexpected low overall multiplicity seen in Au+Au collisions points
to a strong degree of coherence compatible with the onset of saturation in the 
wave function of the ions. A saturation that appears below a transverse momentum scale whose value 
increases with the energy of the collisions and the atomic number (A=197) of the colliding nuclei. 

Asymmetric reactions like d+Au run in collider mode are fertile ground for QCD studies 
 because the projectile and target
fragmentation regions are well separated, and the detection of particles in the
fragmentation regions skews even more the kinematics at the 
partonic level. (See appendix A). These asymmetric systems are thus ideal to study
the small-x components of the  Au target wave function. BRAHMS, one of the RHIC experiments specially
designed to study particle production at high rapidity, was thus particularly well suited to produce
measurements that are considered a first indication of saturation in the gluon density of the Au ion.

\section{Lower energy p+A measurements}
Before proceeding with the description of BRAHMS studies of d+Au collisions, 
a brief review of previous measurements is necessary to emphasize the
novelty of the results that we are going to present.
Early studies of p+A collisions were conducted with the primary aim of 
extracting the proton energy loss in nuclear matter. All those
measurements were done in fixed target mode. I base this review in two well
know papers \cite{Fredriksson, Busza}. Compared with the RHIC measurements, 
all these experiments suffer from the fact that in fixed target mode the 
projectile fragmentation region (close to the proton rapidity) is boosted to
very small angles and a detailed transverse momentum dependent studies were
just not possible.

Many of these measurements have concentrated in particle production as function
of polar angle or pseudo-rapidity $\eta = -log(tan\frac{\theta}{2})$. 

\begin{figure}[!ht]
\begin{center}
\resizebox{0.8\textwidth}{!}
           {\includegraphics{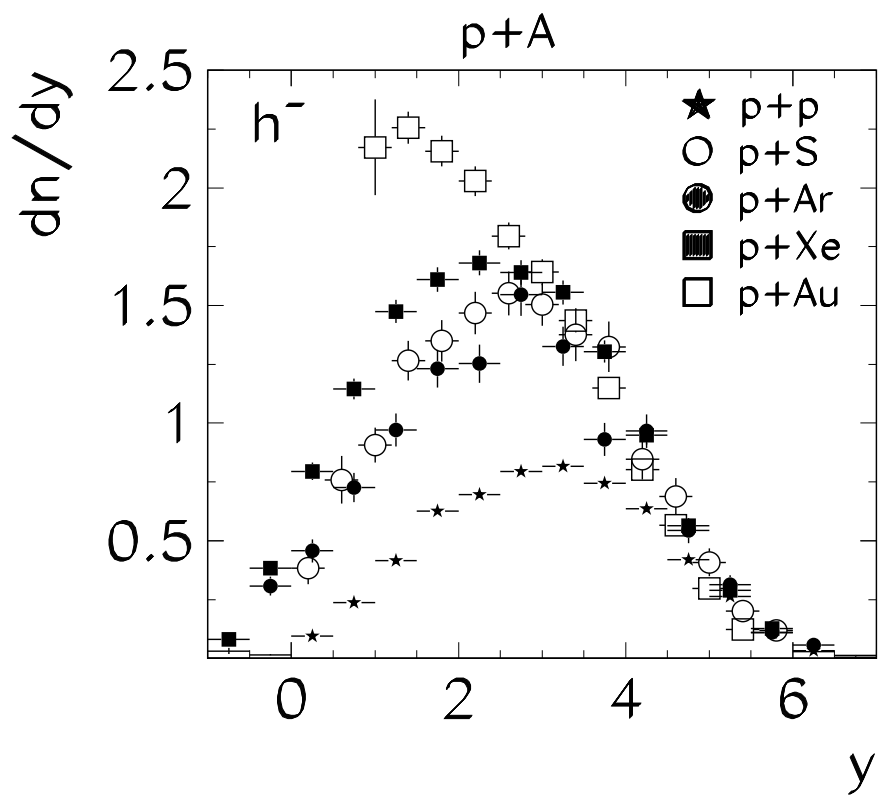}}
\end{center}
\caption{\label{fig:NA35} Rapidity distribution for negative particles 
(mostly negative pions) produced in p+p  and p+A collisions at $\sqrt{s_{NN}} = 19.4 GeV$ at the SPS \cite{NA35}}.

\end{figure}

Figure \ref{fig:NA35} is a compilation of the NA35 pA program at CERN in the form of multiplicity distributions in 
laboratory rapidity space. These distributions have two main features: the first being the fact that most of the yield (in this particular 
case, negative particles or mostly pions) 
from p+A collisions appears close to the target fragmentation regions (y or $\eta \sim 0$ in the lab. reference frame) 
and the second is the fact that within $\sim 1$ unit of rapidity close to the beam rapidity, the projectile proton has no 
more ``memory'' of the target. Similar behavior has been shown to be independent of the beam energy (one such 
compilation in the projectile reference frame can be found in ref. \cite{Busza}).
If the yields from p+A collisions are compared to those from p+p at the same energy, one obtains a
characteristic wedge like distribution that has been explained in the context of multiple parton 
interactions; the projectile appears as if it had only one interaction but that interaction can involve several partons
from the target nucleons. The ratio starts at $\eta = 0$ with a value equal to the numbers of partonic interactions
 and ends at the rapidity of the projectile with a value equal to one. 
 \cite{Brodsky}.

\begin{figure}[!ht]
\begin{center}
\resizebox{0.8\textwidth}{!}
           {\includegraphics{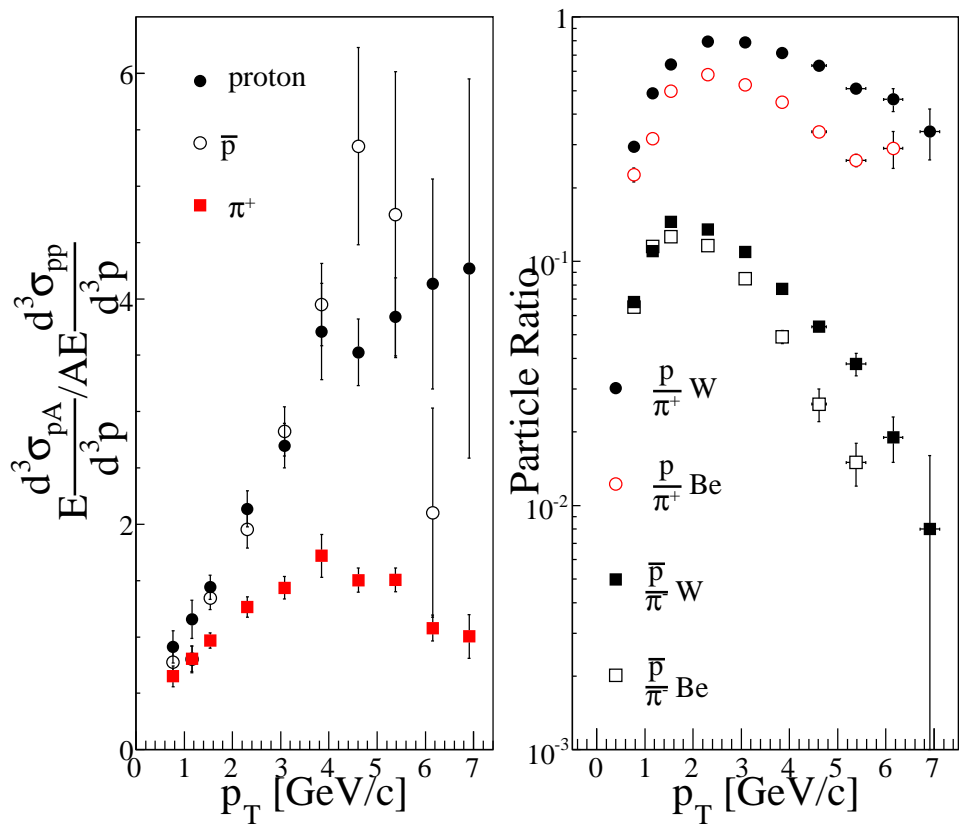}}
\end{center}
\caption{\label{fig:Antresyan} Left panel: Nuclear modification factor for pions, protons and anti-protons at 
mid-rapidity in p+W at $\sqrt{s_{NN}} = 27.4 GeV$ at Fermilab \cite{Antresyan}. These data were collected with 
the spectrometer at a fixed angle, protons and anti-protons at low $p_{T}$ have  rapidities smaller but still close 
to mid-rapidity. Right panel: Ratios of
baryon to mesons for heavy (W) and light (Be) targets measured around mid-rapidity in the same experiment.}

\end{figure}

The measurements of hadrons at large transverse momentum done at Fermilab \cite{Antresyan} have shown 
what is now called the Cronin effect; an enhancement that is widely considered as incoherent multiple elastic 
interactions as the projectile moves
through the target, each one of these interactions modifies the transverse momentum distributions by 
shifting counts from low $p_{T}$
values up to intermediate values ($\sim 4-5 GeV/c$). The nuclear modification factor defined as a ratio of
differential cross sections normalized by the atomic number A of the targets is used to compare 
to an incoherent sum of p+p collisions at the same 
energy. One such comparison is shown in the left panel of Fig. \ref{fig:Antresyan}. The ratio for pions has a 
clear enhancement 
that starts above 2  and extends to 7 GeV/c. Anti-protons show an strikinly different behavior when 
compared to the above described pions.  The difference between baryon and meson present at this energy 
($\sqrt{s_{NN}} = 27.4\ GeV$) has also been seen at RHIC. The right panel of the figure shows a comparison of 
abundances of baryons 
(protons and anti-protons)
and mesons (pions). The ratio of anti-proton to negative pions is small and consistent 
with hadronization in the vacuum, the ratio of protons to positive pions, is greater and approaches one for
the heaviest target. Because the energy of these collisions is not that high, it may be,  that this ratio is 
affected by beam
protons, as some degree of stopping is expected to transfer protons to mid-rapidity where this ratio was constructed.

\begin{figure}[!ht]
\begin{center}
\resizebox{0.8\textwidth}{!}
           {\includegraphics{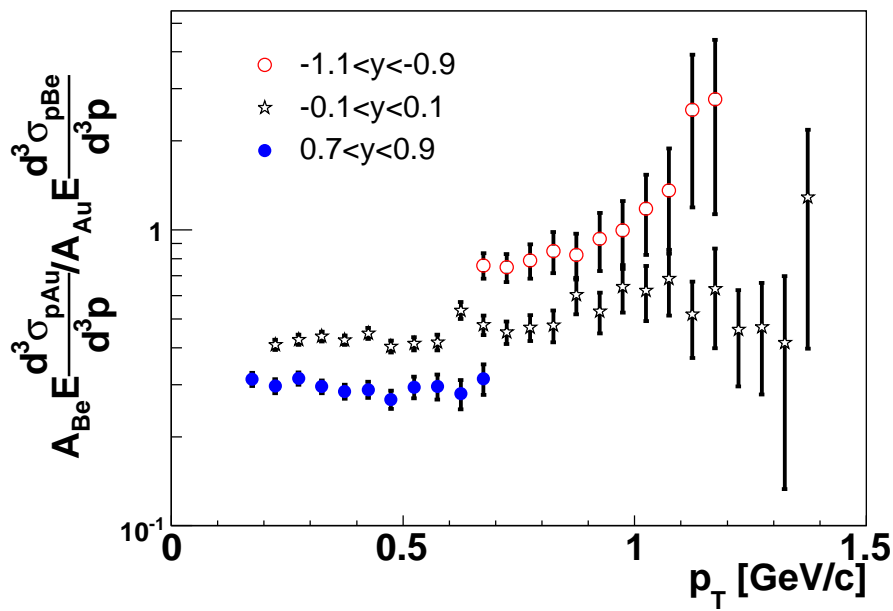}}
\end{center}
\caption{\label{fig:E802} Nuclear modification factor contructed from invariant cross sections 
for positive pions produced in p+Au and p+Be collisions at $\sqrt{s_{NN}} = 5 GeV$ at the AGS ($y_{CM} = 1.7$) in
three rapidity intervals \cite{E802}.}

\end{figure}

Figure \ref{fig:E802} shows results from collisions at an even lower energy. This time, the nuclear modification 
factor is defined as a ratio of invariant 
differential cross sections for positive pion production, scaled by the atomic number of the targets.
The comparison is done between a heavy target (Au) and a light one (Be). The data was collected in fixed target
mode at the AGS with the E802 spectrometer.
The large acceptance coverage in rapidity provides some access to both target and projectile fragmentations regions. 
The rapidity of the 14.6 GeV/c proton beam is equal to 3.4 and the data points in the figure are labeled 
with the rapidities in the nucleon-nucleon center of mass ($y_{CM} = 1.7$).

The most striking feature of this figure is the fact that the curves are arranged in the same 
descending order as the above mentioned ``triangular distribution''. Only the ratio calculated close to target 
rapidities ($-1.1<y<-0.9$) crosses the value of 1 at $p_T \sim 1 GeV/c$. The ratios corresponding to mid-rapidity
$y = 0$ and $y \sim 1$ have values smaller than one. According to the scattering models 
used to explain the Cronin effect the low momentum (ratio below 1) is depleted because each rescattering shifts
the event to higher $p_T$ bins. The depletion is stronger as the rapidity increases because the reach into lower
values of x in the target wave function is greater; more scattering centers are thus available. 
Naive scattering models imply that at higher rapidities, the Cronin peak appears at higher values of $p_T$,
however, data extending to high $p_{T}$ values is not available. It should also be said that available phase space limits 
the applicability of such arguments.
  
\section{Intermediate $p_{T}$ studies and the nuclear modification factors}
BRAHMS is one of the four RHIC experiments with the unique capability to measure identified 
hadrons with transverse momenta that can reach moderately high values ($\sim 5 GeV/c$) and can 
access rapidities close to the beam rapidity (y=5.4 for the 100 GeV/c per nucleon beam). The data that is 
described in this presentation is thus a first detailed study of particle production in the beam fragmentation
region in d+Au collisions at the highest energy in the center of mass. As such, the data may be a window to
new phenomena and in particular, it has been listed as a first indication that the small-x components of the 
target wave function have entered a non-linear mode. 
The data collected from d+Au collisions is compared to p+p using the so called nuclear 
modification factor defined as: $R_{dAu}=\frac{1}{N_{coll}}\frac{\frac{dN^{dAu}}{dp_{T}d\eta}}{\frac{dN^{pp}}{dp_{T}d\eta}}$. If the target is already a saturated system of gluons, the ratio is expected to show a decrease in value as the rapidity of the detected particles increases. If the target is a dilute system of gluons, the ratio should grow with rapidity because, at higher rapidities, the detected particle is related to a parton that has interacted with a greater 
number of small-x gluons, each contributing a finite amount of transverse momentum, such that the ratio beyond some
value of $p_T$ grows greater than one and then tends to one from above.  

\begin{figure}[!ht]
\begin{center}
\resizebox{0.8\textwidth}{!}
           {\includegraphics{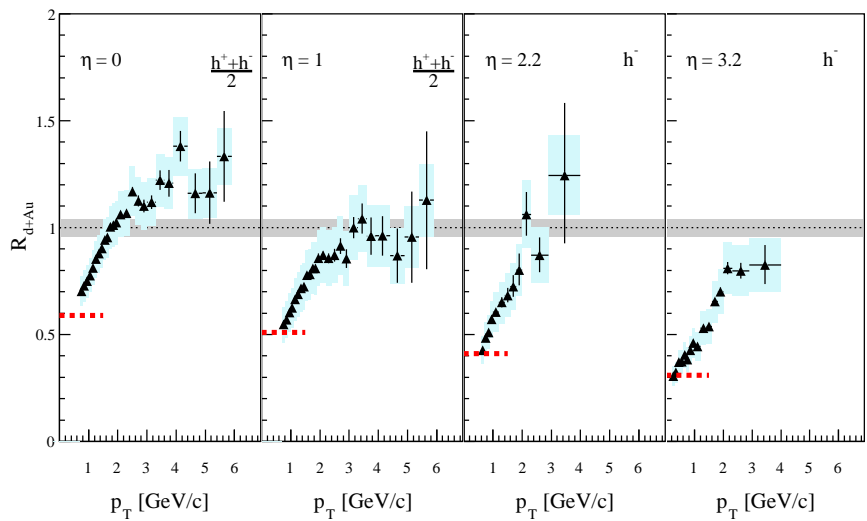}}
\end{center}
\caption{\label{fig:ratio} Nuclear modification factor for charged
  hadrons at pseudorapidities $\eta=0,1.0,2.2,3.2$. Statistical
  errors are shown with error bars. Systematic
  errors are shown with shaded boxes 
  with widths set by the bin sizes.                                          
  The 
  shaded band around
  unity indicates the estimated error on the normalization to $\langle N_{coll} \rangle$. 
  Dashed lines at $p_T<1$ GeV/c show the normalized charged particle 
  density ratio $\frac{1}{\langle
  N_{coll}\rangle}\frac{dN/d\eta(d+Au)}{dN/d\eta(pp)}$.}
\end{figure}

Figure \ref{fig:ratio} shows the nuclear modification factor
with  the number of binary collisions set to $N_{coll} = 7.2 \pm 0.6$ for
 minimum biased d+Au 
collisions.  This particular study was done without identifiying the particles. Each panel 
 shows the ratio calculated at
a different pseudo-rapidity $\eta$ values. 
At mid-rapidity ($\eta = 0$), the nuclear modification factor exceeds 1 for
transverse momenta greater than 2 GeV/c in similar way as the pions in Fig. \ref{fig:Antresyan}. 
 
One unit of rapidity towards the deuteron rapidity is enough to make the  enhancement disappear, 
and then  become consistently smaller than 1 for the next two values of pseudo-rapidity 
($\eta$ = 2.2 and 3.2) 
indicating a suppression in d+Au collisions compared to scaled p+p systems at the same energy. 

The novelty of this result stands mainly on the fact that the measurement extends to moderate 
transverse momenta ($\sim 3.5 GeV/c$) and the factor appears consistently suppressed for 
all value of $p_{T}$, specially at $\eta = 3.2$.

\begin{figure}[!ht]
\begin{center}
\resizebox{0.8\textwidth}{!}
           {\includegraphics{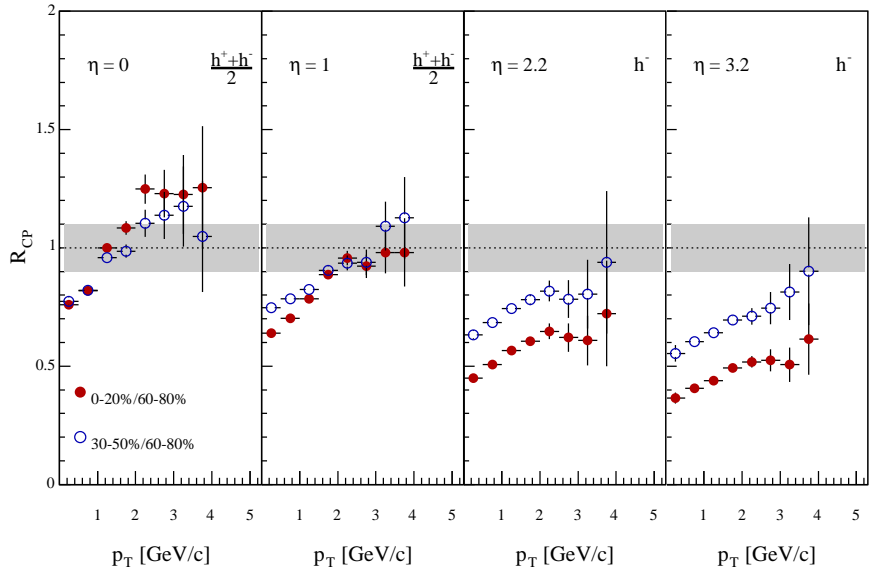}}
\end{center}
\caption{\label{fig:centrality} Central (full points) and
    semi-central (open points) $R_{cp}$ ratios (see text for details)
    at pseudorapidities $\eta=0,1.0,2.2,3.2$. Systematic errors ($\sim5\%$) 
    are smaller than the symbols. The ratios at the highest pseudorapidities ($\eta=2.2$ and 3.2) are calculated for 
negative hadrons. The uncertainty on the normalization of the ratios is displayed as a shaded band 
around unity. Its value has been set equal to the error in the calculation of $N_{coll}$ in the most peripheral collisions ($12\%$). }
\end{figure}

The four panels of Fig. \ref{fig:centrality} show the central $R^{central}_{CP}$ (filled symbols) and semi-central $R^{s
emi-central}_{CP}$
(open symbols) ratios for the four $\eta$ settings. Central events have a higher number of target nucleons 
participating in the interaction with the deuteron projectile. This higher number of nucleons translates into
an increased number of gluons present in the system and, if the conditions are set for saturation, central
collisions would have stronger suppression as function of rapidity. If the target is still a linear dilute 
system the  $R^{central}_{CP}$ would be enhanced as fuction of rapidity because of the higher number of 
scattering centers that become available in the target.    
In the left panel of Fig. \ref{fig:centrality} corresponding to $\eta=0$, the 
 yield from the central sample of events (filled symbols) is 
systematically higher than those of the semi-central events, but at the highest pseudo-rapidity  $\eta=3.2$, 
the trend is reversed; the yields from central events are 
$\sim 60\%$ lower than the semi-central events at all values of $p_{T}$.  More details on these results can be found in \cite{rdAPRL}.
These results have been described within the context of the Color Glass Condensate \cite{McLerranVenu}; 
the evolution of the nuclear modification factor with rapidity and centrality is consistent with a description of the 
Au target where  the rate of gluon fusion  
becomes  comparable with that of gluon emmission as the rapidity increases and it slows down the overall growth of the gluon
density. The measured nuclear modification factor compares the slowed down growth of the numerator to a sum of incoherent 
p+p collisions, considered as dilute systems, whose gluon densities grow faster with rapidity because of the abscence of
gluon fusion in dilute systems \cite{KKTandOthers}. Other explanations for the measured suppression
have been proposed and they also reproduce the data \cite{RudyHwa, Vitev, Kopeliovich}. 

\begin{figure}[!ht]
\begin{center}
\resizebox{0.8\textwidth}{!}
           {\includegraphics{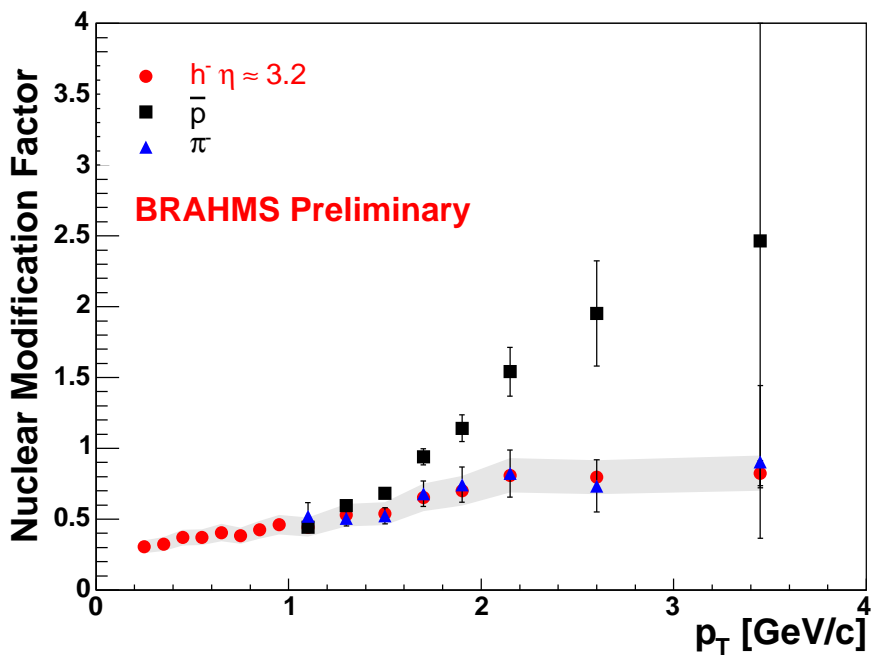}}
\end{center}
\caption{\label{fig:identifiedRdA}The nuclear modification factor $R_{dAu}$ calculated for
anti-protons (filled squares) and negative pions (filled triangles) at $\eta = 3.2$. The same 
ratio calcutated for negative particles at the same pseudo-rapidity \cite{rdAPRL} is shown with filled circles, 
and the systematic error for that
measurement is shown as grey band. }
\end{figure}

The nuclear modification factor of baryons is different from the one calculated with mesons, whenever the factor shows
the so called Cronin enhancement, baryons show a stronger enhancement. Such difference has been seen ot lower energies and 
is shown for pions and anti-protons in Fig. 
\ref{fig:Antresyan}, it has also been found at RHIC energies at all rapidities, in particular,  Fig. \ref{fig:identifiedRdA} presents the minimum bias nuclear modification $R_{dAu}$ for anti-protons and negative pions at $\eta = 3.2$. These ratios were obtained making use of ratios of raw counts of identified particles compared to
those of charged particles in each $p_{T}$ bin: 

\[ R^{\bar{p}}_{dAu} = 
R^{h^-}_{dAu} \frac{(\frac{\bar{p}}{h^-})^{dAu}}{(\frac{\bar{p}}{h^-})^{pp} } = 
\frac{1}{N_{coll}}\frac{\left.\frac{dn^{dAu}}{dp_{T}d\eta}\right)^{h^{-}}}{\left.\frac{dn^{pp}}{dp_{T}d\eta}\right)^{h^{-}}} \frac{\frac{\left.\frac{dn^{dAu}}{dp_{T}d\eta}\right)^{\bar{p}}}{\left.\frac{dn^{dAu}}{dp_{T}d\eta}\right)^{h^{-}}}}
{\frac{\left.\frac{dn^{pp}}{dp_{T}d\eta}\right)^{\bar{p}}}{\left.\frac{dn^{pp}}{dp_{T}d\eta}\right)^{h^{-}}}} = 
\frac{1}{N_{coll}}\frac{\left.\frac{dn^{dAu}}{dp_{T}d\eta}\right)^{\bar{p}}}{\left.\frac{dn^{pp}}{dp_{T}d\eta}\right)^{\bar{p}}} \]

 No attempt was made to estimate the contributions from anti-lambda feed down 
to the 
anti-proton result. The remarkable difference between 
baryons and mesons has  been related to parton recombination
\cite{RudyHwa}.  

\begin{figure}[!ht]
\begin{center}
\resizebox{0.8\textwidth}{!}
           {\includegraphics{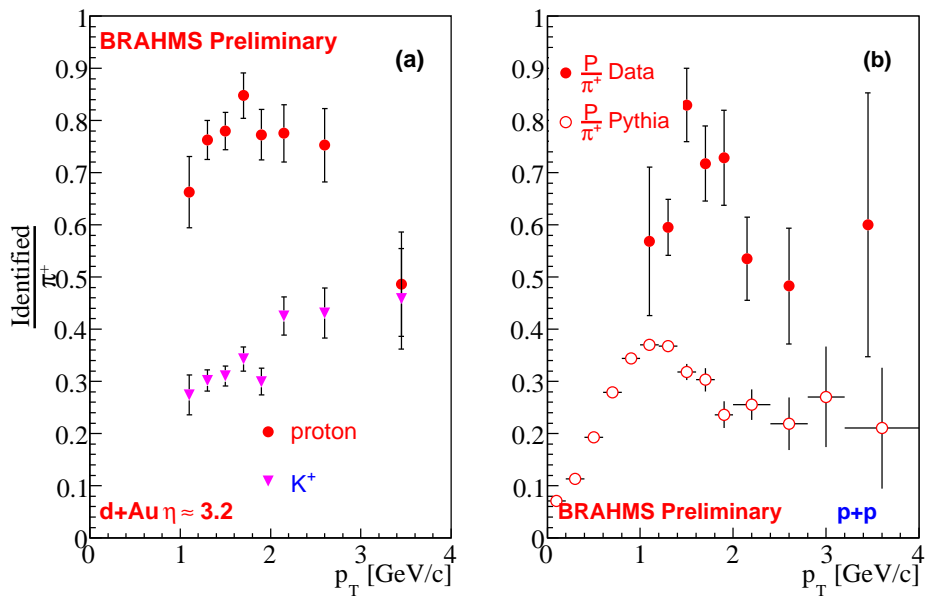}}
\end{center}
\caption{\label{fig:identifiedRatiosdA}The fraction of proton over positive pions at $\eta = 3.2$: (a) Particle 
composition of positive charged hadrons produced in d+Au collisions at $\eta=3.2$ The abundance of protons and kaons 
are compared to the one of pions as function of transverse momentum. (b) The same comparison but this time for 
particles produced at the same rapidity and energy in p+p collisions. }
\end{figure}

The ratios shown in Fig. \ref{fig:identifiedRatiosdA} show a new aspect of particle production at forward rapidities.
The left panel shows that in d+Au collisions at $\eta = 3.2$ the yield of protons is comparable to the pions yield ($\sim 80\%$) while the 
yield of positive kaons hovers around $40\%$. These results indicate that eventhough pion production is well described
at all rapidities in p+p collisions \cite{Bland}, the presence of so many baryons at that rapidity brings additional 
complications to NLO pQCD calculations, which cannot be reconciled with the data  
if standard fragmentation functions are used \cite{Strickman}.  The abundance of baryons at this high energy and rapidity 
doesn't support the idea of baryon 
suppression in the fragmentation region \cite{Dumitru, Strikman2} where, because of their high energy, the quarks 
of the beam would fragment independently mostly into mesons. But if that suppression was actually present much closer to beam
rapidity, baryon number conservation would force the transfer of beam protons to lower rapidities, what remains a mystery is
the mechanism that gives these protons the high transverse momentum ($\sim 2 GeV/c$) that is measured. Panel b of
Fig. \ref{fig:identifiedRatiosdA} shows a similar baryon excess in p+p collisions at the same high rapidity. This time
the comparison is made to Pythia simulations.

In summary, particle production  from d+Au and p+p collisions at $\sqrt{s_{NN}}= 200 GeV$ and 
at different rapidities with the BRAHMS setup offers a window to the small-x 
components of the Au wave function. The suppression found in the particle production at high rapidities from d+Au 
collisions may be the first 
indication of the onset of saturation in the gluon distribution function of the Au target. 

\section{Acknowledgments}

This work was supported by 
the Office of Nuclear Physics of the U.S. Department of Energy, 
the Danish Natural Science Research Council, 
the Research Council of Norway, 
the Polish State Committee for Scientific Research (KBN) 
and the Romanian Ministry of Research.

\appendix
\section{Kinematics of pA}

The Parton 
model describes hadrons moving at very high momentum as combinations of systems of massles partons in 
numbers that grow as the energy of the probe increases. Each parton carries
a fraction x of the hadron momemtum P. (Because of the very high momentum of the hadron any transverse motion
can be neglected.) 

\begin{figure}[!ht]
\begin{center}
\resizebox{0.8\textwidth}{!}
           {\includegraphics{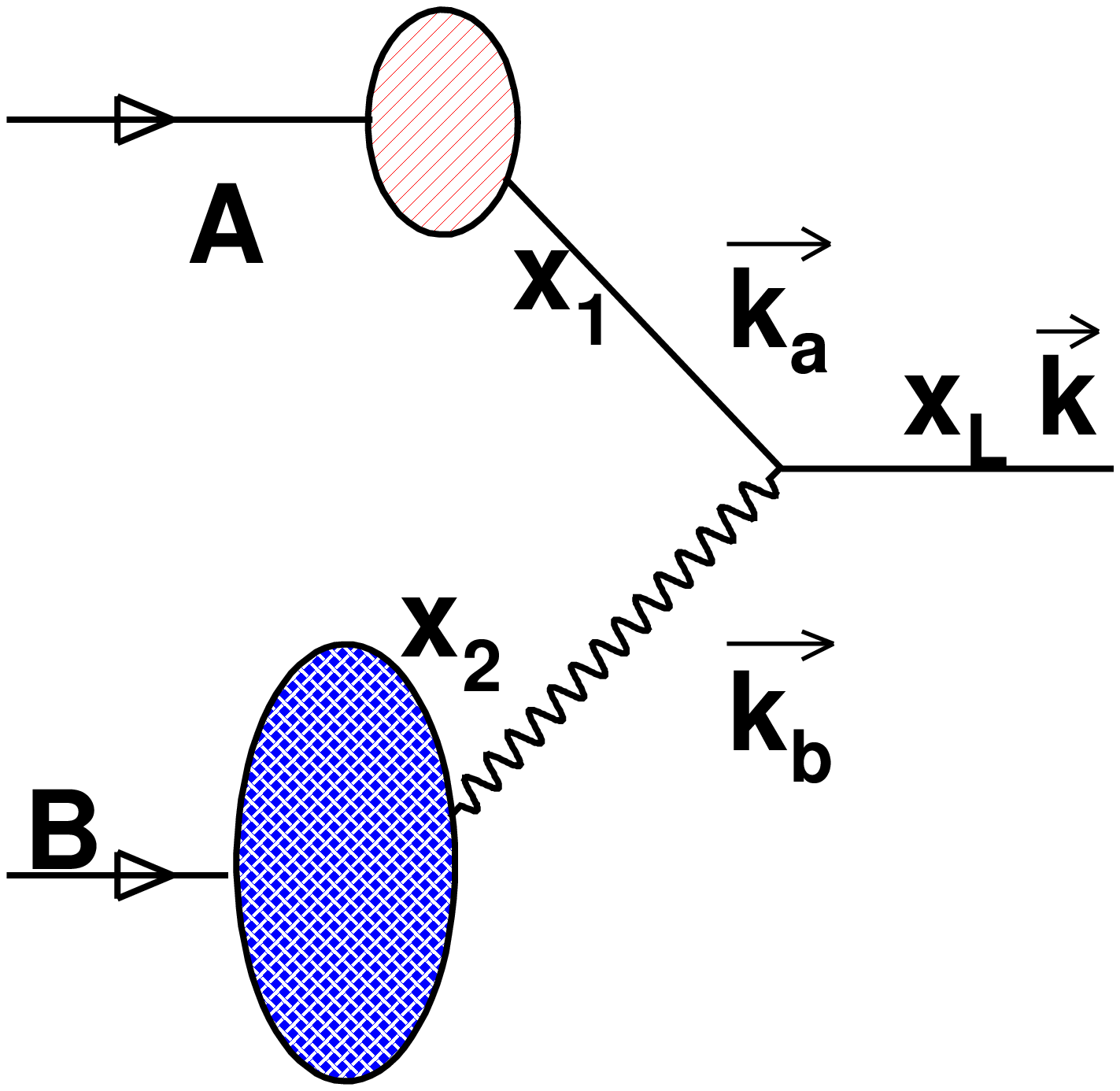}}
\end{center}
\caption{\label{fig:dAdiagram} Diagram representing a $2\rightarrow 1$ process in a collision 
between two nuclei A and B (A would be the deuteron ion and B the gold ion).
The detected particle has longitudinal momentum fraction $x_{L}$ and rapidity y. }
\end{figure}

What follows is the derivation of the connection between the longitudinal fractions $x_1$ and $x_2$  of the two partons 
in the initial state and the rapidity and transverse mass of the measured particle as well as the overall energy of the 
collision.
This derivation is done for the simpler 2->1 case but similar results are obtained for the more general 2->2 interactions.
These derivations were done with much help from Chelis Chasman.

Let S be the total center-of-mass energy squared: $S = (\bf{P}_A + \bf{P}_B)^{2}$ where 
$\bf{P}_{A}$ and $\bf{P}_{B}$ are four vectors. (Naturally the beam momentum defines one prefered direction, 
and from now 
on we state that the four-momenta $\bf{P}_{A}$ and $\bf{P}_{B}$ have only one component along that direction. If the 
momenta of the colliding nucleons
is high compared to their masses, one can neglect the masses and write:

$$S = (E_A + E_B)^2 - (\vec{P_A} + \vec{P_B})^2 = m_{A}^2 + m_{B}^2 +2E_AE_B - 2\vec{P_A}\cdot \vec{P_B} \equiv 2 \bf{P}_A \cdot \bf{P}_B $$
The masses of the nucleons can be neglected (E = P for both nuclei and $P_A = -P_B$) and one writes in the center of mass of the collision:
$$S = 4P_A P_B $$

The fraction x of the hadron's longitudinal  momentum carried by each parton is 
defined as the
ratio of the appropriate light-cone momenta:
$x \equiv \frac{e_{parton} + p_{parton}^{\parallel}}{E_{hadron} + P_{hadron}}$ for a left-to-right moving hadron and  
$x \equiv \frac{e_{parton} - p_{parton}^{\parallel}}{E_{hadron} - P_{hadron}}$ for the right-to-left hadron.
Using the same labels shown in Fig. A1 we call $x_1$ the fraction of longitudinal momentum of a beam 
(proton or deuteron) parton:
$$x_1 = \frac{e_b +  p_{b}^{\parallel}}{E_B + P_B} = \frac{e_b +  p_{b}^{\parallel}}{\sqrt{S}} $$
where we neglected the mass of the hadrons and wrote: $E = P = \frac{\sqrt{S}}{2}$.

For the parton in the heavy nuclei (Au) we write the fraction $x_2$  emphasizing the fact that the hadron now moves from 
right-to-left i.e. $\vec{P_A}<0$ :
$$x_{2} = \frac{e_a -  p_{a}^{\parallel}}{E_A - P_A} = \frac{e_a +  p_{a}^{\parallel}}{\sqrt{S}} $$

$$x_{1}\sqrt{S} = e_{b} +  p_{b}^{\parallel} \mbox{ and } x_{2}\sqrt{S} = e_a +  p_{a}^{\parallel}$$

$$x_{1}\sqrt{S}(e_b -  p_{b}^{\parallel}) = m_{b}^2 + k_{b}^2 = ( m^{b}_{\perp})^{2} \mbox{ and } x_{2}\sqrt{S}(e_a -  p_{a}^{\parallel}) = m_{a}^2 + k_{a}^2 =( m^{a}_{\perp} ) ^{2}$$

where $k_{a}$ and $k_{b}$ are the transverse momentum components of the interacting partons and  $ m_{\perp}^{a}$ and 
$ m_{\perp}^{b}$ are called transverse masses of the a and b partons respectively. If we 
multiply the squares of those transverse masses we get:

$$  (m^{a}_{\perp})^{2}  (m^{b}_{\perp})^{2} = Sx_{1}x_{2}(2e_ae_b + 2p_{a}^{\parallel} p_{b}^{\parallel} - (e_{a} + p_{a}^{\parallel})(e_{b} + p_{b}^{\parallel})) $$

where we used the fact that the partons move in opposite directions along the longitudinal axis.
 
If we define $\hat{s}$ as total energy in the center-of-mass of the interaction partons a and b:

$$\hat{s} = (e_a + e_b)^2 -  (p_{a}^{\parallel} + p_{b}^{\parallel})^2 - (k_{a} + k_{b})^2 = M^2 $$

where M is the mass of the new system formed by the ineraction of the partons a and b. That system has transverse 
motion given by: $\vec{p_T} = \vec{k_{a}} + \vec{k_{b}}$  and we can write $\hat{s}$ in the following way:

$$\hat{s} = (m^{a}_{\perp})^2 + (m^{b}_{\perp})^2 + 2e_ae_b + 2p_{a}^{\parallel} p_{b}^{\parallel} - p_{T}^2  $$
$$ 2e_ae_b + 2p_{a}^{\parallel} p_{b}^{\parallel} = \hat{s} - (m^a_{\perp})^2 - (m^{b}_{\perp})^2 +  p_{T}^2  $$

when we replace the left side of this equation in the product of transverse masses derived a few lines above, we have:

$$\hat{s} - (m^a_{\perp})^2 - (m^{b}_{\perp})^2 +  p_{T}^2 = \frac{ (m^{a}_{\perp})^{2}  (m^{b}_{\perp})^{2}}{ Sx_{1}x_{2}} $$
$$\hat{s} +   p_{T}^2 = M_{T}^{2} =  x_{1}x_{2}S + \frac{ (m^{a}_{\perp})^{2}  (m^{b}_{\perp})^{2}}{ Sx_{1}x_{2}} + (m^a_{\perp})^2 +  (m^{b}_{\perp})^2$$

If one neglects the masses of the partons and their transverse momentum motion: 

$$ x_{1}x_{2}S = M_{T}^2$$ 

On the other hand, if the system formed at the interaction of partons a and b has rapidity y and longitudinal 
momentum $p_L$,  its longitudinal 
momentum fraction is defined in the center-of-mass frame as: $x_{L} = \frac{2 p_L}{\sqrt{S}}$ 
longitudinal momentum conservation for the 2 to 1 process is written as:

$$x_1 -x_2 = x_{L} = \frac{2 M_T}{\sqrt{s}} sinh y$$
 
and together with the relation between energy in the nucleon-nucleon centrer of mass S and the one in the parton-parton system, we have a system of two equations with two unknowns $x_1$ and $x_2$.

$$x_1 = \frac{M_T}{\sqrt{S}} e^y$$
$$x_2 = \frac{M_T}{\sqrt{S}} e^{-y}$$

It is now clear how the work at high rapidities opens a window into the low values of $x_2$ in the target wave function.
\end{document}